\documentclass[a4paper,11pt]{article}
\usepackage{pos}
\usepackage{cleveref}
\usepackage{hyperref}

\title{Scattering from generalised lattice $\phi^4$ theory}

\author*[a]{Marco Garofalo
}
\author[b,c]{Fernando Romero-L\'opez}
\author[a,d]{Akaki Rusetsky}
\author[a]{Carsten Urbach}

\affiliation[a]{HISKP (Theory), Rheinische Friedrich-Wilhelms-Universit\"at Bonn, Nussallee 14-16, 53115 Bonn, Germany}
\affiliation[b]{Center for Theoretical Physics, Massachusetts Institute of Technology,Cambridge, MA 02139, USA}

\affiliation[c]{IFIC, CSIC-Universitat de Val\`encia, 46980 Paterna, Spain}
\affiliation[d]{Tbilisi State University, 0186 Tbilisi, Georgia}

\emailAdd{garofalo@hiskp.uni-bonn.de}

\abstract{We investigate numerically different techniques to extract scattering amplitudes from the Euclidean Lattice $\phi^4$ theory with two fields, having different masses. We present an exploratory study of the recently proposed method by Bruno and Hansen for extracting the scattering length from a four-point function \cite{Bruno:2020kyl}, and a study of the two- and three-particle quantization condition.}

\FullConference{%
 The 38th International Symposium on Lattice Field Theory, LATTICE2021
  26th-30th July, 2021
  Zoom/Gather@Massachusetts Institute of Technology
  
  Report number: MIT-CTP/5366
}

\graphicspath{{plots/}}

\begin{document}
\maketitle

\section{Introduction}

The standard approach to determine scattering quantities from Lattice QCD
 is the Lüscher method~\cite{Luscher:1990ck}, which relates the finite-volume spectrum obtained from the lattice to the infinite-volume scattering amplitude.
  It has been applied to many physical systems, see Ref.~\cite{Briceno:2017max} for a review. The formalism has also been recently extended to three particles with three different but conceptually equivalent formulations available in the literature at present~\cite{Hansen:2014eka,Hansen:2015zga,Hammer:2017uqm,Hammer:2017kms,Mai:2017bge}, see Refs.~\cite{Hansen:2019nir,Mai:2021lwb} for recent reviews.
  
In this contribution we study the techniques to extract scattering amplitudes
from the Euclidean Lattice field theory. We use a Euclidean Lattice $\phi^4$ theory with two fields having different masses. Using this theory has proven to be an excellent test environment for novel scattering studies, as shown in Refs.~\cite{Sharpe:2017jej,Romero-Lopez:2018rcb,Romero-Lopez:2020rdq}.

In particular we study the recent proposal \cite{Bruno:2020kyl}, in which the authors found a relation
between  the scattering length and the Euclidean four-point functions at threshold kinematic. Henceforth, this will be referred to as the BH method.
We compare the BH method to the standard Lüscher approach and find good agreement.
This study of the BH method presented here is based on \cite{Garofalo:2021bzl}.

We also investigate the extraction of the scattering quantities at non-zero momentum. In particular we studied s-wave scattering amplitude for two particles with the Lüscher method~\cite{LUSCHER1991531}.

\section{Description of the Model}

The Euclidean model used here is composed of two real scalar fields $\phi_i, i=0,1$ with the Lagrangian
  \begin{align}
  {\cal L}= \sum_{i=0,1} \left( \frac{1}{2} \partial_\mu \phi_i\partial_\mu \phi_i +\frac{1}{2}m_i \phi_i^2 +\lambda_i \phi_i^4\right)
+\mu \phi_0^2 \phi_1^2\, ,
\label{eq:lagrangian} 
 \end{align}
with nondegenerate (bare) masses $m_0<m_1$. The Lagrangian has a $Z_2\otimes Z_2$ symmetry $\phi_0\to-\phi_0$ and $\phi_1\to-\phi_1$, which prevents sectors with even and odd number of particles to mix. 

To study the problem numerically,
we define the theory on a finite hypercubic lattice with lattice spacing $a$ and a volume $T \cdot L^3$, where $T$ denotes the Euclidean time length and $L$ the spatial length. We define the derivatives of the Lagrangian (\ref{eq:lagrangian}) on a finite lattice as the finite differences $\partial_\mu \phi(x)=\frac{1}{a}\,(\phi(x+a\mu)-\phi(x))$.  In addition, periodic boundary conditions are assumed in all directions. The discrete action is given in Ref.~\cite{Romero-Lopez:2018rcb} for the complex scalar theory, but it is trivial to adapt it to this case. We set $a=1$ in the following for convenience.



\section{The BH method}

In Ref.~\cite{Bruno:2020kyl}, Bruno and Hansen derived a relation between the scattering length $a_0$ and the following combination of Euclidean four-point and two-point correlation functions at the two-particle threshold:
\begin{equation}
    C_4^\mathrm{BH}(t_f,t,t_i) \equiv   \frac{\langle \tilde\phi_0(t_f,0)\tilde\phi_1(t,0)\tilde\phi_1(t_i,0) \tilde\phi_0(0,0)\rangle}
{\langle \tilde\phi_0(t_f,0) \tilde\phi_0(0,0)\rangle \langle \tilde\phi_1(t,0)\tilde\phi_1(t_i,0) \rangle} -1,
\end{equation}
with the time ordering $t_f > t > t_i > 0$,  and  $\tilde \phi_{i}(t,{\bf p})=\sum_{\bf x}e^{ i \bf p\cdot \bf x }\phi_{i}(t,\bf x)$ being
spatial Fourier transform   of the field. In particular $\tilde \phi_{i}(t,0)$ is the field  projected to zero spatial momentum. The relation of $C_4^\mathrm{BH}$ to the scattering length reads 
\begin{align}
    C_4^\mathrm{BH}(t_f,t,t_i)
    \xrightarrow[t\gg t_i\gg 0]{T\gg t_f\gg t}
    \frac{2}{ L^3}&\bigg[ \pi \frac{a_0 }{\mu_{01} }(t-t_i)
      - 2a_0^2\sqrt{\frac{2 (t-t_i)}{\mu_{01}} }+ O\left((t-t_i)^0\right)\bigg]\,,
    \label{eq:BH}  
\end{align}
where $\mu_{01}=(M_0 M_1)/(M_0+M_1)$ is the reduced mass. It is defined in terms of the renormalized masses $M_0$ and $M_1$ of the two particles. These masses can be extracted as usual from an exponential fit at large time distances of the two-point correlation functions 
\begin{equation}
\langle \tilde\phi_{i}(t,{\bf p})\tilde\phi_{i}(0,-{\bf p}) \rangle \approx
   A_{1,i} \left(e^{ - E^{i}_1({\bf p})  t}  + e^{ - E_1^{i}({\bf p}) (T - t)}\right)
   \label{eq:2pt}
\end{equation}
  with  $E_1^{i}({\bf p} =0)=M_{i}$ for $i=0,1$. To reduce the statistical error we average over all points with the same source sink separation.

\subsection{Numerical result}

We generate ensembles using the Metropolis-Hastings algorithm with the bare masses $m_0=-4.925$ and $m_1=-4.85$, and for simplicity we choose $\lambda_0=\lambda_1=2\mu=2.5$.  The list of ensembles generated in this work with their corresponding measured values of the masses $M_0$ and $M_1$ are compiled in \cref{tab:full_table}. In this model,
two-point correlators are dominated by the ground state from the first time slice. This was also observed
 in previous investigations of the scalar theory \cite{Romero-Lopez:2020rdq}.

\begin{figure}
    \centering
    \includegraphics[scale=0.48]{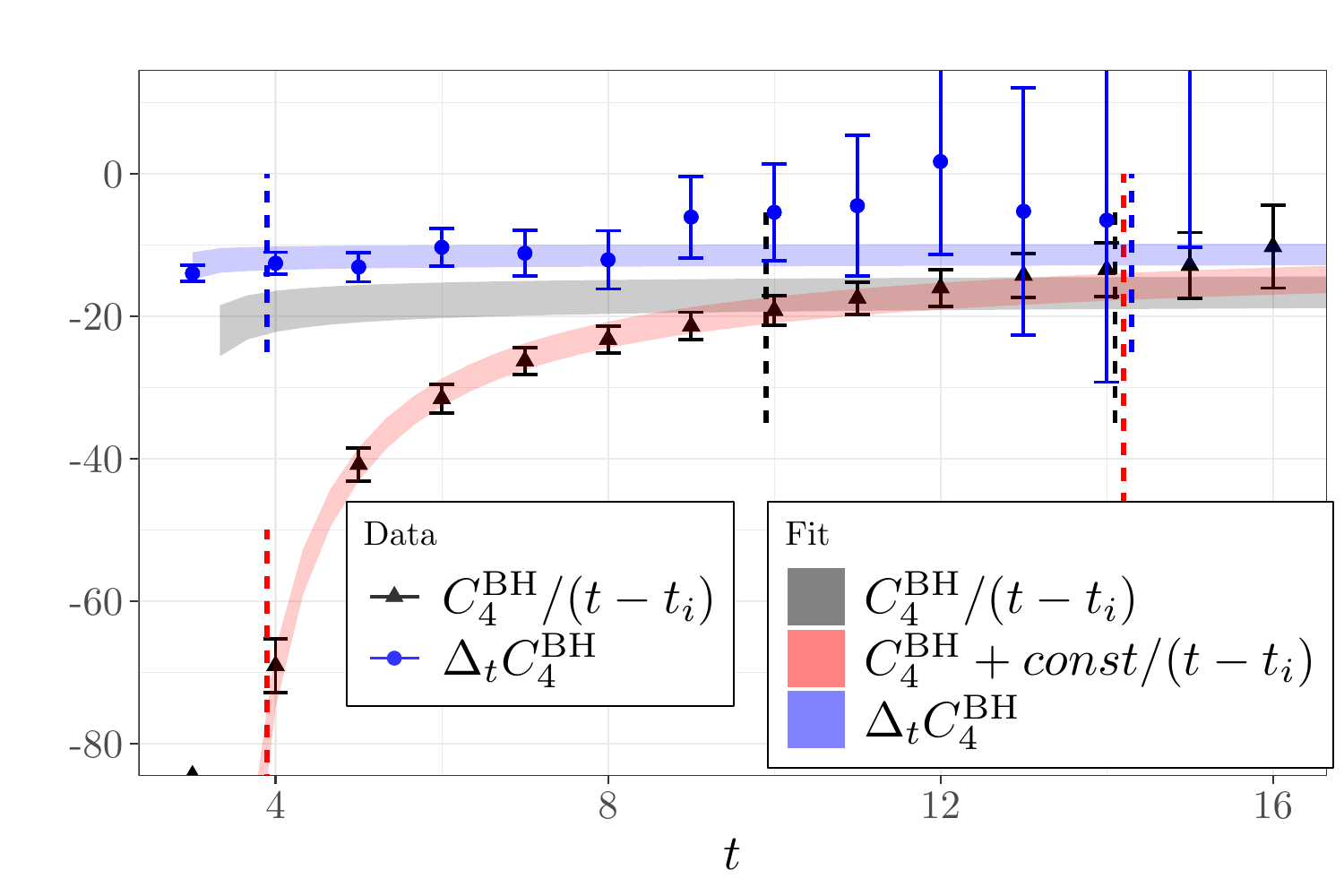}
    \includegraphics[scale=0.48]{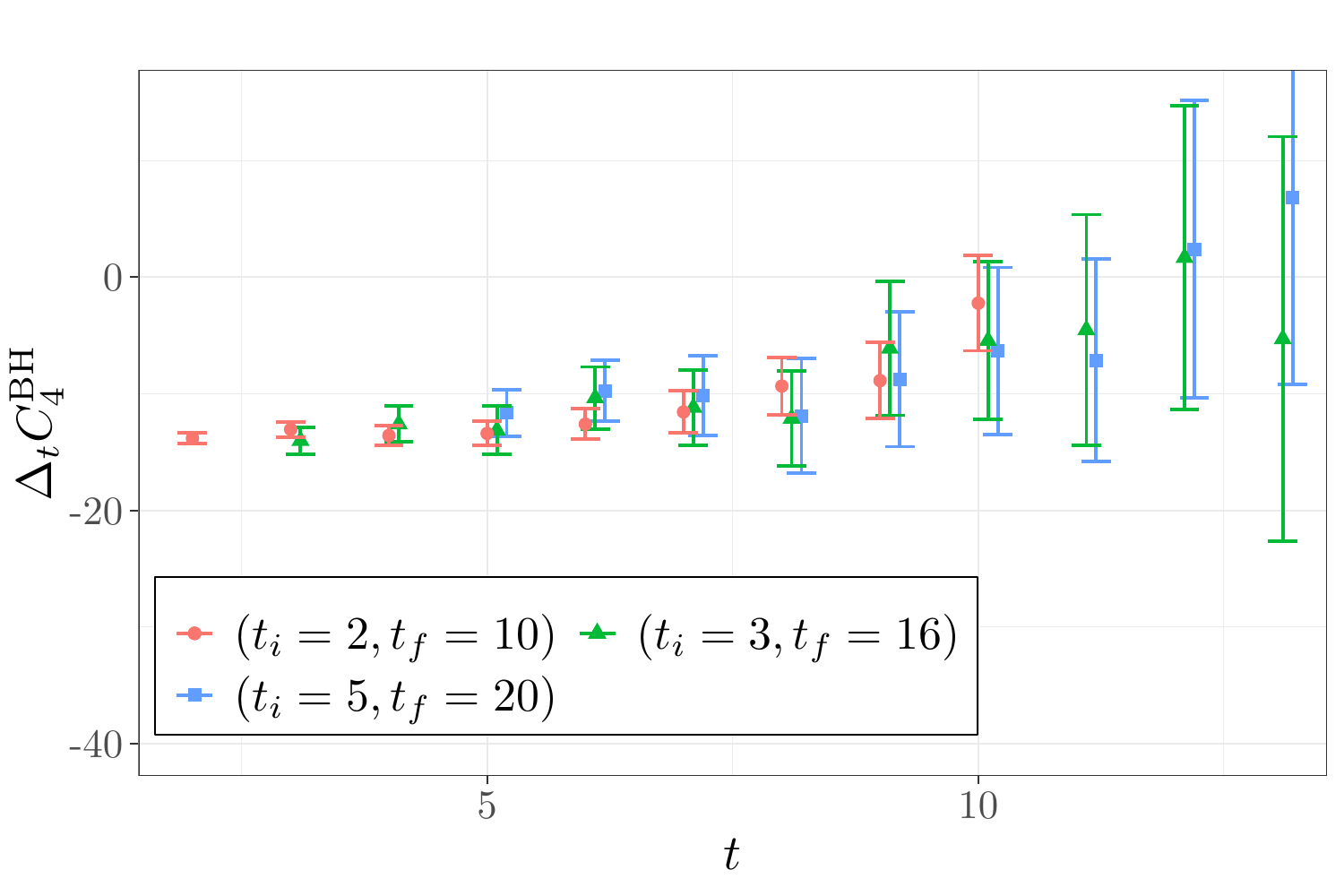}
    \caption{Left panel: Four-point function of \cref{eq:BH} multiplied by $L^3/2$, for $L=22$ and $T=96$ with $t_i=3$ and $t_f=16$ divided by $(t-t_i)$ (black triangles). the dashed vertical lines represent the fit interval, the black band represent the result of the fit \cref{eq:BH} and the red band is the same fit with an extra constant term. The blue circles and band represent the discrete derivative of the correlator \cref{eq:Delta_BH} and the corresponding  fit.
Right panel: Plot of the discrete derivative of the correlator \cref{eq:Delta_BH}
    for different values of $t_i$ and $t_f$. We do not observe any systematic shift and all correlators 
    are compatible. The points with smaller $t_i$ and $t_f$
    tend to have smaller error.    
    }
    \label{fig:BH_03t16}
\end{figure}

We tried three different strategies to extract the scattering length:
\begin{enumerate}
\item We attempt a direct fit of \cref{eq:BH} the the data. 
\item We include an overall constant in the fit to account for the $O\left((t-t_i)^0\right)$ effect. 
\item We make use of the shifted function at fixed $t_i$ and $t_f$,
$\Delta_t C_4^\mathrm{BH}(t_f,t,t_i)= C_4^\mathrm{BH}(t_f,t+1,t_i)-C_4^\mathrm{BH}(t_f,t,t_i)$,
where the constant term cancels out. We then determine $a_0$ by fitting to
  \begin{equation}
    \label{eq:Delta_BH}
    \begin{split}
      \Delta_t C_4^\mathrm{BH}(t_f,t,t_i&) \approx \frac{2}{L^3}\Big[\pi\frac{a_0}{\mu_{01}}\Big. 
        \Big.-2a_0^2 \sqrt{\frac{2}{\mu_{01}}}\Big(\sqrt{t+1-t_i} - \sqrt{t-t_i}\,\Big)\Big].
    \end{split}
  \end{equation}
\end{enumerate}
The three methods are compared in the left panel of \cref{fig:BH_03t16} for one of our ensembles. The black triangles represent the correlator of \cref{eq:BH} with $t_i=3$ and $t_f=16$ divided by $(t-t_i)$ so that  it goes to a constant when $(t-t_i) \to \infty$.
The monotonic increase of the data points could be due to the $\left((t-t_i)^0\right)$ term in \cref{eq:BH}. A fit to the formula \cref{eq:BH} in the time region $[10,14]$---the black band--- gives a good $\chi^2/d.o.f\sim0.7$, but results in large uncertainties. The quality of the fit deteriorates very quickly if the fit range is extended: a fit in the time region $[6,14]$ yields $\chi^2/\mathrm{dof}\sim 5$.

With the second strategy---the red band in the left panel of \cref{fig:BH_03t16}---one is able to start fitting at significantly smaller $t$-values.
The data are well described with a $\chi^2/\mathrm{dof}\sim 0.2$ 

For the third approach, we study $\Delta_t C_4^\mathrm{BH}(t)$. This is shown in the left panel of \cref{fig:BH_03t16} as blue circles, and the blue band represents the best fit result with error.
The main advantage of the latter strategy is that it allows us to extract the physical information at smaller $t$ without introducing extra parameters in the fit. 
Indeed, the data looks almost constant over the complete $t$-range available.
Only very close to $t_i$ the square root term might become visible.

For the third strategy, which looks most promising from a systematic point of view, we also investigate the dependence on the choice of $t_{i}$ and $t_{f}$. This is shown in right panel of \cref{fig:BH_03t16} for the same ensemble as in the left panel.
We do not observe any significant systematic effect stemming from excited state contributions when changing $t_i$ or $t_f$.
However, we clearly see significantly smaller statistical uncertainties with smaller $t_i$ and $t_f$ values.

\subsection{Comparison to the Lüscher method}

\begin{figure}
    \centering
    \includegraphics[scale=0.6]{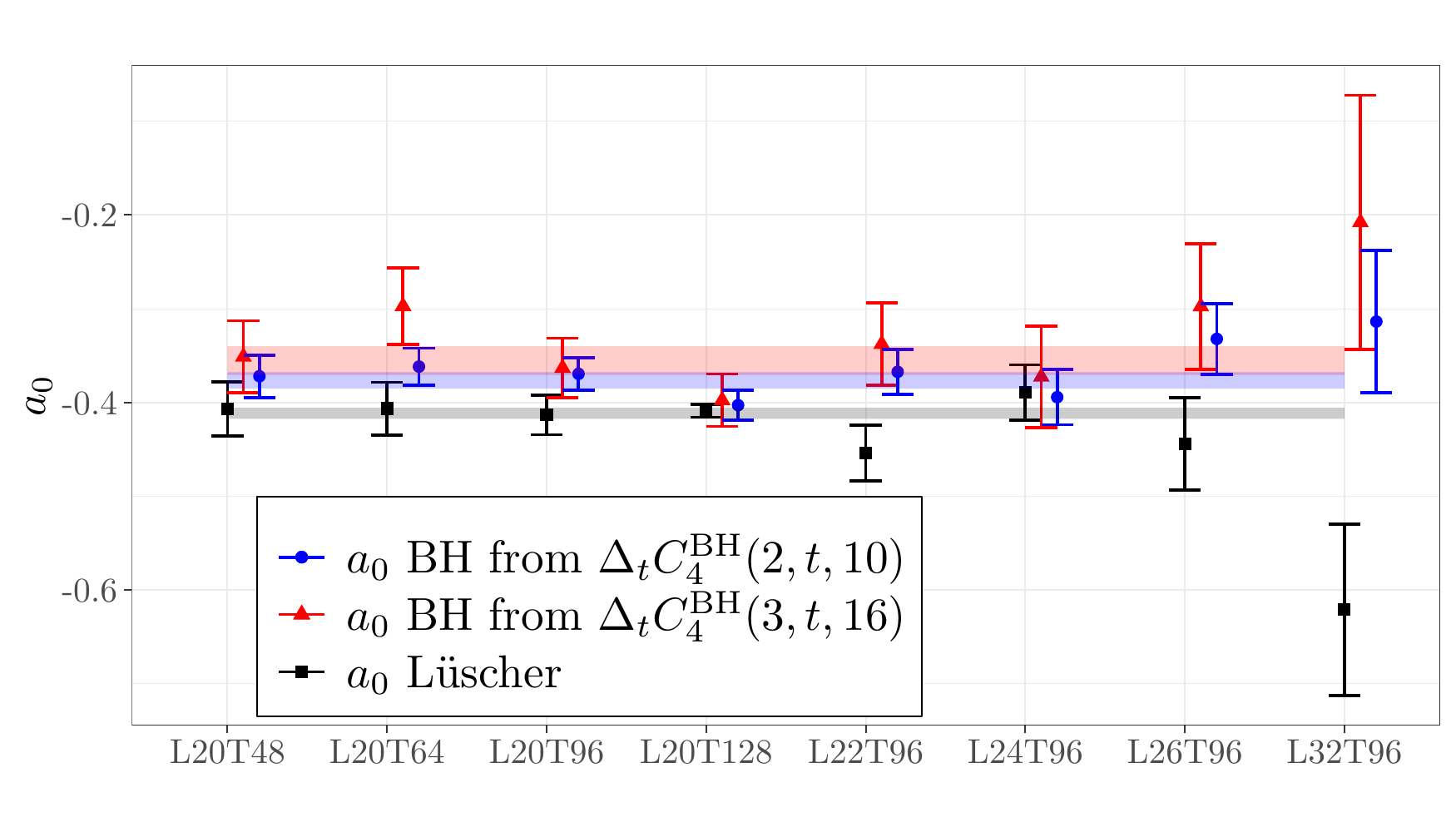}
    \caption{ Comparison of $a_0$ computed with  BH method \cref{eq:Delta_BH} with $t_i=2$ and $t_f=10$ (blue circles), with $t_i=3$ and $t_f=16$ (red triangles) and  Lüscher method  \cref{eq:luescher_a0} (black squares) the horizontal bands correspond to the weighted average of each method.}
    \label{fig:compare_L_BH}
\end{figure}

In this section we compare the BH method described above with the Lüscher threshold expansion~\cite{Luscher:1985dn,Beane_2005}. The latter relates the two-particle energy shift, defined as $\Delta E_{2}^{01}=E_{2}^{01}-M_0-M_1$, to the scattering length $a_0$ via
\begin{equation}
  \Delta E_{2}^{01} =-\frac{2\pi a_{0}}{\mu_{01} L^3}\left[ 1 + c_1  \frac{a_{0}}{L} + c_2\left(\frac{a_{0}}{L}\right)^2  \right] +O\left(L^{-6}\right)\,,
  \label{eq:luescher_a0}
\end{equation}
with $c_1=-2.837297$, $c_2= 6.375183$ and $E_2^{01}$ being the interacting two-particle energy at zero total momentum.
$E_2^{01}$ can be extracted from 
$C_2(t) = \langle \tilde\phi_1(t,0)\tilde\phi_0(t,0)  \tilde\phi_1(0,0)\tilde\phi_0(0,0) \rangle$, whose large-$t$ behaviour is
\begin{align}
    \begin{split}
         C_2(t)     \xrightarrow[T-t\gg0]{t\gg0}     A_2 e^{-E_2^{01} \frac{T}{2}} \cosh{\left(E_2^{01} (t-\frac{T}{2})\right)} 
     +B_2  e^{-(M_0+M_1) \frac{T}{2}} \cosh{\left((M_1-M_0) (t-\frac{T}{2})\right)}\,.     \label{eq:E2_01}
    \end{split}
\end{align}
with the last term being a thermal pollution due to finite $T$ with periodic boundary condition.
Using $M_0$ and $M_1$ as input determined from the corresponding two-point functions, the only additional parameter is $B_2$.
Alternatively, it is possible to eliminate the second term defining
$
\tilde C_2(t)=C_2(t)/\cosh{\left((M_1-M_0) (t-\frac{T}{2})\right)},
$
and then taking the finite derivative
\begin{align}
    \begin{split}
  \Delta_t\tilde C_2(t)&=\tilde C_2(t+1)-\tilde C_2(t).
    \label{eq:Delta_E2_01}
    \end{split}
\end{align}
The two-particle energies obtained from \cref{eq:E2_01} are compatible with those from \cref{eq:Delta_E2_01}. The results are reported in \cref{tab:full_table}, along with the values for 
 the scattering length $a_0$ computed from $E_2$ using \cref{eq:luescher_a0}. 
 A comparison between the BH and the Lüscher method is depicted in \cref{fig:compare_L_BH} for all our ensembles. The values are compatible with each other. However, the BH method
 gives systematically larger values for $a_0$.
For each ensemble separately Lüscher and BH methods appear compatible.
However we observe a systematic trend after averaging over all ensembles, as shown in the bands of fig.~\ref{fig:compare_L_BH}.
This might be attributed to different lattice artifacts.
The statistical error is similar in both approaches. Also, the scaling in $L$ appears to be similar.
The different systematics of the two methods offer in general a useful opportunity for cross-checks.

\begin{table*}
{\footnotesize
    \centering
    \setlength{\tabcolsep}{3pt}
    
   \setlength{\tabcolsep}{3pt}
    \begin{tabular}{cc|cc|cc|cc|ccc}
\hline\hline
T & L & \(M_0\) & \(M_1\) &\multicolumn{2}{c|}{$E_2^{01}$} &\multicolumn{2}{c|}{$a_0$ Lüscher} & \multicolumn{3}{c}{$a_0$ BH}    \\
\hline
 &  &  &  & \(  C_2\) & \( \Delta_t\tilde C_2\) & \(  C_2\) & \( \Delta_t\tilde C_2\) & \(\Delta_t C_4^\mathrm{BH}\)(3,t,16) 
& \(C_4^\mathrm{BH}+c\)  & \(\Delta_t C_4^\mathrm{BH}\)(2,t,10)  \\
\hline
48 & 20 & 0.14675(5) & 0.27487(5) & 0.4252(3) & 0.4253(3) &
-0.41(3) & -0.42(3) & -0.35(4) & -0.35(6) & -0.37(2) \\
64 & 20 & 0.14659(5) & 0.27480(5) & 0.4249(3) & 0.4250(3) &
-0.41(3) & -0.41(4) & -0.30(4) & -0.29(6) & -0.38(2) \\
96 & 20 & 0.14662(4) & 0.27487(4) & 0.4251(2) & 0.4251(3) &
-0.41(2) & -0.41(3) & -0.36(3) & -0.36(4) & -0.38(1) \\
96 & 22 & 0.14604(3) & 0.27470(4) & 0.4237(2) & 0.4237(3) &
-0.45(3) & -0.45(5) & -0.34(4) & -0.31(6) & -0.37(2) \\
96 & 24 & 0.14574(4) & 0.27458(4) & 0.4223(2) & 0.4221(3) &
-0.39(3) & -0.36(6) & -0.36(5) & -0.41(7) & -0.39(2) \\
96 & 26 & 0.14547(4) & 0.27455(3) & 0.4218(2) & 0.4219(3) &
-0.44(5) & -0.47(8) & -0.30(7) & -0.3(1) & -0.36(3) \\
96 & 32 & 0.14521(4) & 0.27449(4) & 0.4210(2) & 0.4213(3) &
-0.62(9) & -0.7(1) & -0.2(1) & -0.1(2) & -0.35(5) \\
128 & 20 & 0.14668(3) & 0.27484(3) & 0.42509(7) & 0.4251(3) &
-0.409(7) & -0.41(3) & -0.40(3) & -0.39(3) & -0.40(1) \\
\hline\hline
\end{tabular}
    \caption{Measured values of $a_0$, $M_0$, $M_1$ and $E_2$.
    The column \(\Delta_t C^\mathrm{BH}_4\) corresponds to the value of $a_0$ fitted with \cref{eq:Delta_BH} fixing $t_i=3$ and $t_f=16$ or $t_i=2$ and $t_f=10$.
     The column \(C_\mathrm{BH}+c\) is the result of the fit with \cref{eq:BH} adding a constant term. The two-particle energy 
     $E_2$ is computed from $C_2$ with the fit of \cref{eq:E2_01} and from $\Delta \tilde C_2$ with 
     \cref{eq:Delta_E2_01}. The corresponding value of $a_0$ computed with the Lüscher method is reported in the corresponding columns.
     We used $2\cdot10^{7}$ configurations for each ensemble, generated from 200 replicas each of
     $10^5$ thermalized configurations. We bin the configurations in blocks of $10^5$ (the entirely replica) and we resample the resulting 200 configurations with jackknife. For the light mass $M_0$ we measured the integrated autocorrelation time  $\tau_{int}\sim1.5$ , while $\tau_{int}\sim0.5$ for $M_1$. We skip 1000 configurations in each replica for thermalization.
    }
    \label{tab:full_table}
}
\end{table*}

\section{Scattering amplitude at not zero momentum with the L\"uscher method}

\begin{figure}
\includegraphics[scale=0.65,trim=0 0 0 0]{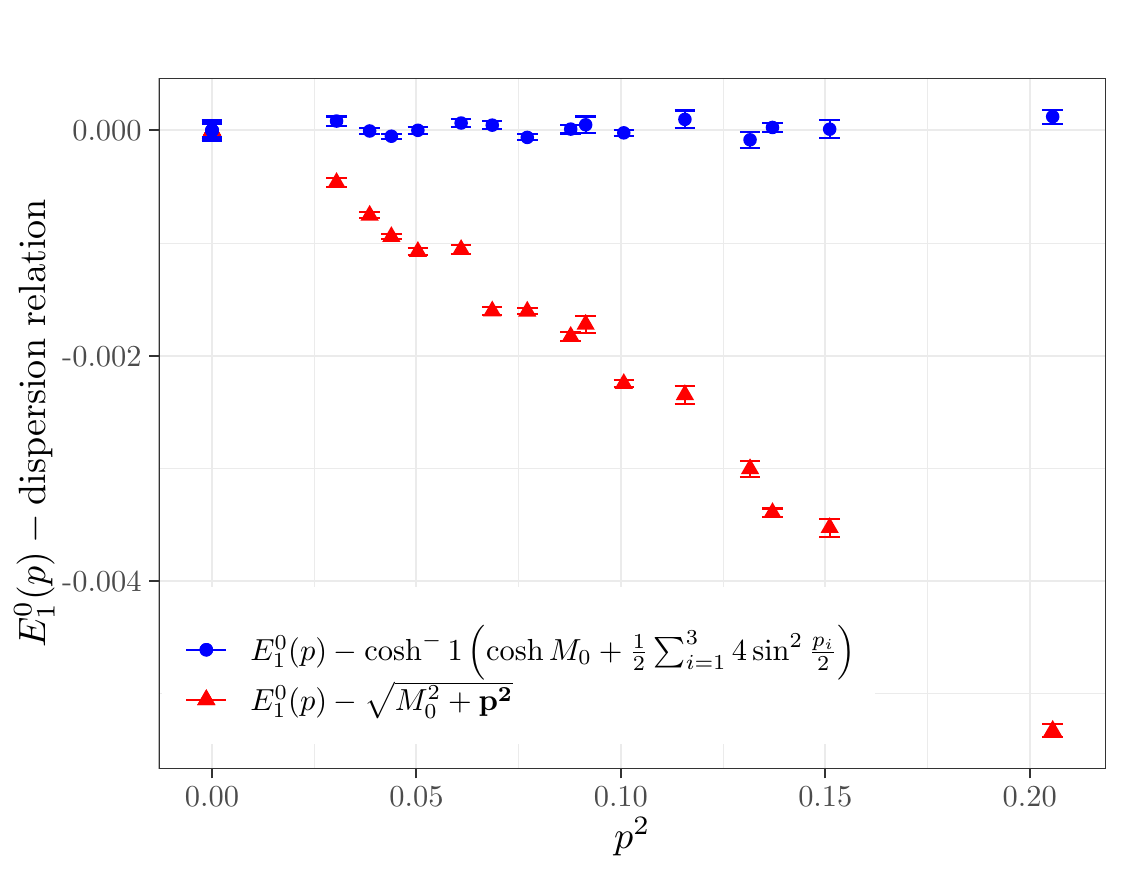}
\includegraphics[scale=0.65]{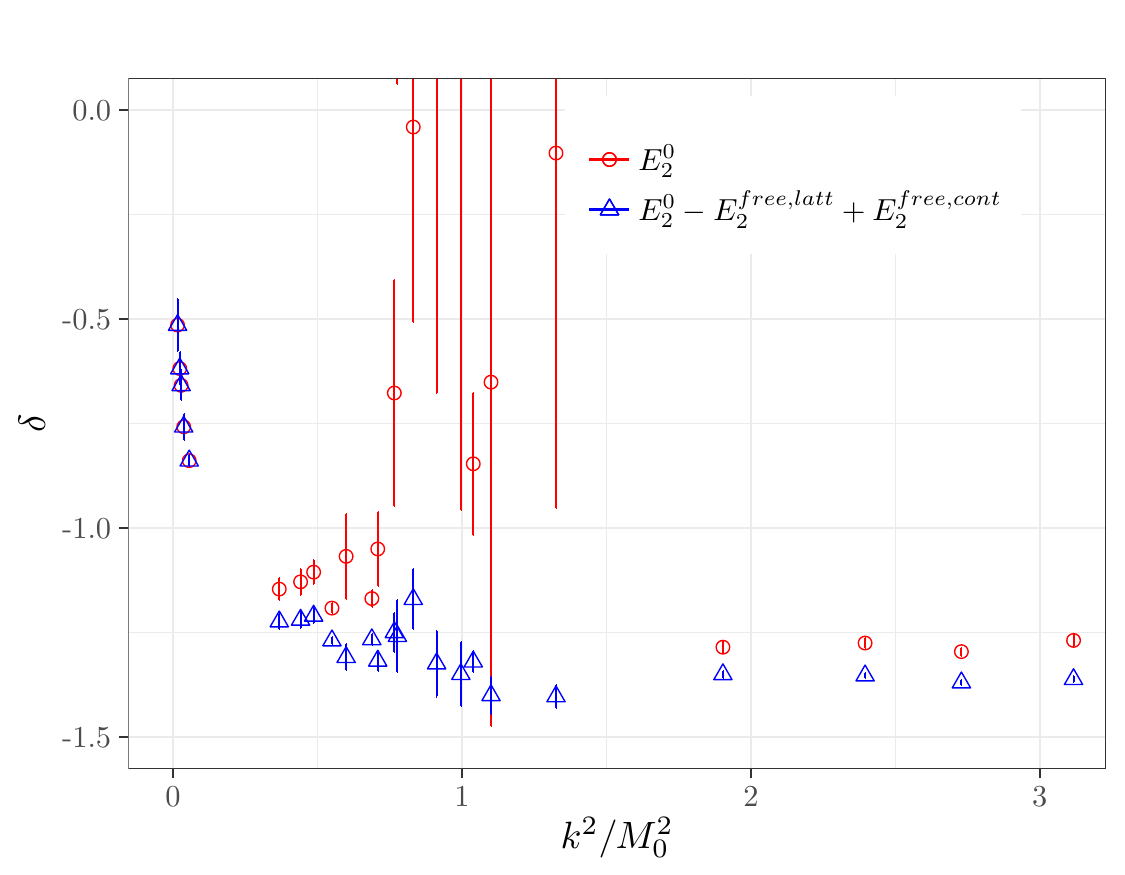}
\caption{
Left panel: Values of the energies level measured for different values of $p^2$ minus the value predicted using the lattice dispersion relation (blue circles) or the continuum dispersion relation (red triangles).
Right panel: values of the phases shift $\delta$ obtained from \cref{eq:luescher_qc} using as input the two-particle energies measured on the lattice (red circles) or the energies corrected by a lattice artifact $E_2^{free,latt}-E_2^{free,cont}$ (blue point).
The statistic used is the same as described in the caption of \cref{tab:full_table}. }
\label{fig:kcot}
\end{figure}

In this section, we report our study of the  s-wave scattering amplitude for two particles with the Lüscher method \cite{LUSCHER1991531}.
We compute the spectrum of our $\phi^4$ model at ${\bf p}\neq 0$ for the lighter
particle. We generate ensembles using the Metropolis-Hastings algorithm with bare masses $m_0=-4.9$ and $m_1=-4.65$, to have $M_1\sim 3M_0$, keeping  $\lambda_0=\lambda_1=2\mu=2.5$.
We extract the energies from an exponential fit to \cref{eq:2pt}
with the discretised momentum ${\bf p} = 2\pi {\bf n}/L$ with ${\bf n}$ being a integer vector in the set  ${\bf n}\in \{ (0, 0, 0), (1, 0, 0), (1, 1, 0), (1, 1, 1)\}$.
As we can see from the left panel of  \cref{fig:kcot}, we find that the measured one-particle energies  $E_1^0({\bf p})$
significantly deviate from the continuum dispersion relation, while they
 are in good agreement with the lattice dispersion relation
\begin{gather}
   \cosh\left( E_1^0({\bf p})\right)=  \cosh(M_0) +\frac{1}{2}\left( \sum_{i=1}^{3}4 \sin\left(\frac{ p_{i}}{2}\right)^2\right)\,,
    \label{eq:lat-disp-rel}
\end{gather}
with $M_0$ being the mass measured in the fitted in \cref{eq:2pt} at zero momentum.

For each choice of the momentum we construct the two-particle operator in the A1 irrep of the cubic group  $\hat O_2(t,\bf p)=\tilde\phi_0(t,\bf p)\tilde\phi_0(t, 0)$. Only for the first unit of momentum we construct the operator with back to back momentum still in the $A_1$ irrep $\hat O_2(t,0)=\sum_{i=x,y,z}\tilde\phi_0(t, p_i)\tilde\phi_0(t,-p_i)$ with $p_x=(2\pi/L,0,0)$, $p_y=(0,2\pi/L,0)$ and $p_z=(0,0,2\pi/L)$. We measured the two-particle energy from the exponential fit of the correlator
\begin{align}
        \langle \hat O_2 (t,{\bf p}) \hat O_2 (0,-{\bf p}) \rangle\xrightarrow[T-t\gg0]{t\gg0}   &  A_2 e^{-E_2^0 ({\bf p})\frac{T}{2}} \cosh{\left(E_2^0({\bf p}) (t-\frac{T}{2})\right)} \\
     +&A_1  e^{-(E_1^0({\bf p})+M_0) \frac{T}{2}} \cosh{\left((E_1^0({\bf p})-M_0) (t-\frac{T}{2})\right)}\,,
\end{align}
where $E_1^0({\bf p})$ and $M_0$ are the ones obtained from the fit of \cref{eq:2pt}.
From the two-particle energies $E_2^0({\bf p})$ we calculate the s-wave phase shift as \cite{LUSCHER1991531}
    \begin{gather}
             \cot \delta = \frac{ Z_{0,0}(1,q^2) }{\pi^{3/2} \gamma q}
             \,,\label{eq:luescher_qc}
 \end{gather}
 where  $Z_{0,0}$  is the L\"uscher zeta function, the Lorentz boost $\gamma=E_2^0({\bf p})/E_{CM}$ is defined in terms of center of mass energy $E_{CM}=E_2^0({\bf p})-{\bf p}^2$ and $q=kL/2\pi$ with the scattering momentum  $k=\frac{E_{CM}}{4}-M_0^2 $.
We notice that, at large momentum $\bf p$, the values of the phase shift computed with \cref{eq:luescher_qc}  come with large errors (red circles of \cref{fig:kcot}).
These large errors stem from the fact that the energies deviate significantly from the continuum dispersion relation. This problem was also observed in \cite{RUMMUKAINEN1995397}, where as a solution the author propose to compute the center of mass energy $E_{CM}$ using the lattice dispersion relation. 
Here we follow a similar strategy. 
We subtract the difference between  the free two -particle energies 
$E_2^{free,latt}-E_2^{free,cont}$  computed with the lattice \cref{eq:lat-disp-rel} and continuum dispersion relation from the 
two-particle energy.
 The factor $E_2^{free,latt}-E_2^{free,cont}$ is a lattice artifact, i.e. it goes to zero in the continuum limit. However, at finite lattice spacing, we notice a reduction of the statistical error in the phase shift computed from energies at large momentum $\bf$ \cref{fig:kcot} (blue point of the right panel of \cref{fig:kcot}).

\section{Conclusion}
 
 In this contribution, we studied  the BH method,
proposed in \cite{Bruno:2020kyl}. We have verified that it produces
results that are compatible with those of the Lüscher
method \citep{Luscher:1985dn}.
We also studied the s-wave scattering amplitude for two particles with the Lüscher method  at non-zero momentum \cite{LUSCHER1991531}. As in \cite{RUMMUKAINEN1995397}, we observed that the error on the computed phase shift becomes larger  with the momentum of the two-particle state. This increase of the error can be mitigated by the subtraction of a lattice artifact in the measured energy.


\section{Acknowledgements}
 We gratefully acknowledge helpful discussions with M.~Bruno, M.~T.~Hansen
  and S.~R.~Sharpe.
  FRL acknowledges financial support from Generalitat Valenciana grants PROMETEO/2019/083 and CIDEGENT/2019/040, the EU project H2020-MSCA-ITN-2019//860881-HIDDeN, and the Spanish project FPA2017-85985-P.
  The work of FRL is supported in part by the U.S. Department of Energy, Office of Science, Office of Nuclear Physics, under grant Contract Numbers DE-SC0011090 and DE-SC0021006.
  This work
  is supported in part by the Deutsche Forschungsgemeinschaft (DFG,
  German Research Foundation) and the 
  NSFC through the funds provided to the Sino-German
  Collaborative Research Center CRC 110 “Symmetries
  and the Emergence of Structure in QCD” (DFG Project-ID 196253076 - TRR 110, NSFC Grant No. 12070131001).
  AR acknowledges support from Volkswagenstiftung (Grant No. 93562) and the Chinese Academy of Sciences (CAS) President’s International Fellowship Initiative (PIFI) (Grant No. 2021VMB0007).
  The C\texttt{++} Performance Portability Programming Model Kokkos \cite{kokkos} and
  the open source software packages R~\cite{R:2019} have 
  been used.
  We thank B.~Kostrzewa for useful discussions on Kokkos.

\bibliographystyle{unsrt} 
\bibliography{bibliography}

\begin{thebibliography}{10}

\bibitem{Bruno:2020kyl}
Mattia Bruno and Maxwell~T. Hansen.
\newblock {Variations on the Maiani-Testa approach and the inverse problem}.
\newblock {\em Journal of High Energy Physics}, 2021(6), Jun 2021.

\bibitem{Luscher:1990ck}
Martin Lüscher and Ulli Wolff.
\newblock {How to Calculate the Elastic Scattering Matrix in Two-dimensional
  Quantum Field Theories by Numerical Simulation}.
\newblock {\em Nucl. Phys.}, B339:222--252, 1990.

\bibitem{Briceno:2017max}
Raul~A. Briceno, Jozef~J. Dudek, and Ross~D. Young.
\newblock {Scattering processes and resonances from lattice QCD}.
\newblock {\em Rev. Mod. Phys.}, 90(2):025001, 2018.

\bibitem{Hansen:2014eka}
Maxwell~T. Hansen and Stephen~R. Sharpe.
\newblock {Relativistic, model-independent, three-particle quantization
  condition}.
\newblock {\em Phys. Rev. D}, 90(11):116003, 2014.

\bibitem{Hansen:2015zga}
Maxwell~T. Hansen and Stephen~R. Sharpe.
\newblock {Expressing the three-particle finite-volume spectrum in terms of the
  three-to-three scattering amplitude}.
\newblock {\em Phys. Rev. D}, 92(11):114509, 2015.

\bibitem{Hammer:2017uqm}
Hans-Werner Hammer, Jin-Yi Pang, and A.~Rusetsky.
\newblock {Three-particle quantization condition in a finite volume: 1. The
  role of the three-particle force}.
\newblock {\em JHEP}, 09:109, 2017.

\bibitem{Hammer:2017kms}
H.~W. Hammer, J.~Y. Pang, and A.~Rusetsky.
\newblock {Three particle quantization condition in a finite volume: 2. general
  formalism and the analysis of data}.
\newblock {\em JHEP}, 10:115, 2017.

\bibitem{Mai:2017bge}
M.~Mai and M.~D\"oring.
\newblock {Three-body Unitarity in the Finite Volume}.
\newblock {\em Eur. Phys. J. A}, 53(12):240, 2017.

\bibitem{Hansen:2019nir}
Maxwell~T. Hansen and Stephen~R. Sharpe.
\newblock {Lattice QCD and Three-particle Decays of Resonances}.
\newblock {\em Ann. Rev. Nucl. Part. Sci.}, 69:65--107, 2019.

\bibitem{Mai:2021lwb}
Maxim Mai, Michael Döring, and Akaki Rusetsky.
\newblock Multi-particle systems on the lattice and chiral extrapolations: a
  brief review.
\newblock {\em EPJ ST}, 06 2021.

\bibitem{Sharpe:2017jej}
Stephen~R. Sharpe.
\newblock {Testing the threshold expansion for three-particle energies at
  fourth order in $\phi^4$ theory}.
\newblock {\em Phys. Rev. D}, 96(5):054515, 2017.
\newblock [Erratum: Phys.Rev.D 98, 099901 (2018)].

\bibitem{Romero-Lopez:2018rcb}
Fernando Romero-L\'opez, Akaki Rusetsky, and Carsten Urbach.
\newblock {Two- and three-body interactions in $\varphi ^4$ theory from lattice
  simulations}.
\newblock {\em Eur. Phys. J. C}, 78(10):846, 2018.

\bibitem{Romero-Lopez:2020rdq}
Fernando Romero-L\'opez, Akaki Rusetsky, Nikolas Schlage, and Carsten Urbach.
\newblock {Relativistic $N$-particle energy shift in finite volume}.
\newblock {\em JHEP}, 02:060, 2021.

\bibitem{Garofalo:2021bzl}
Marco Garofalo, Fernando Romero-L\'opez, Akaki Rusetsky, and Carsten Urbach.
\newblock {Testing a new method for scattering in finite volume in the $\phi^4$
  theory}.
\newblock 7 2021.

\bibitem{LUSCHER1991531}
Martin Lüscher.
\newblock Two-particle states on a torus and their relation to the scattering
  matrix.
\newblock {\em Nuclear Physics B}, 354(2):531--578, 1991.

\bibitem{Luscher:1985dn}
M.~Lüscher.
\newblock {Volume Dependence of the Energy Spectrum in Massive Quantum Field
  Theories. 1. Stable Particle States}.
\newblock {\em Commun.Math.Phys.}, 104:177, 1986.

\bibitem{Beane_2005}
S.R. Beane, P.F. Bedaque, A.~Parreño, and M.J. Savage.
\newblock A framework for exploring the interactions and decays of hyperons
  with lattice qcd.
\newblock {\em Nuclear Physics A}, 747(1):55–74, Jan 2005.

\bibitem{RUMMUKAINEN1995397}
K.~Rummukainen and Steven Gottlieb.
\newblock Resonance scattering phase shifts on a non-rest-frame lattice.
\newblock {\em Nuclear Physics B}, 450(1):397--436, 1995.

\bibitem{kokkos}
Christian~R. Trott et~al.
\newblock Kokkos 3: Programming model extensions for the exascale era.
\newblock {\em IEEE Transactions on Parallel and Distributed Systems},
  33(4):805--817, 2022.

\bibitem{R:2019}
{R Core Team}.
\newblock {\em R: A Language and Environment for Statistical Computing}.
\newblock R Foundation for Statistical Computing, Vienna, Austria, 2019.

\end{thebibliography}

\end{document}